\documentclass[aps,prb,twocolumn,superscriptaddress]{revtex4} 
\usepackage{dcolumn}%
\usepackage{bm}%
\usepackage{ulem}
\usepackage{setspace}
\usepackage{graphicx}
\usepackage{amsmath}
\usepackage{latexsym}
\usepackage{amsmath}
\usepackage{amssymb}
\usepackage{float}
\usepackage{epstopdf}
\usepackage{hyperref} 
\hypersetup{urlcolor=red,citecolor=blue}
\usepackage{bm}
\newcommand{\sptwo}{$sp^2$}

\newcommand{\pz}{$p_z$}

\newcommand{\spthree}{$sp^3$}
\newcommand{\spthreedfive}{$sp^3d^5$}
\newcommand{\csixty}{C$_{60}$}

\newcommand{\ammonia}{NH$_3$}

\newcommand{\sig}{$\sigma$}
\newcommand{\pai}{$\pi$}
\newcommand{\omeg}{$\Omega$}
\newcommand{\dg}{$^\circ$}
\newcommand{\alfa}{$\alpha$}
\newcommand{\alfabo}{$\alpha_{BO}$}
\newcommand{\alfaomega}{$\alpha_{\Omega}$}
\newcommand{\alfawf}{$\alpha_{WF}$}
\newcommand{\alfat}{$\alpha_t$}
\newcommand{\alfazero}{$\alpha_0$}

\hypersetup{urlcolor=red,citecolor=blue}

\begin{document}
\title{Maximally valent orbitals 
in systems with non-ideal bond-angles} 

\author{Joydev De, Sujith N S, Manoar Hossain, Joydeep Bhattacharjee\\
\small{\textit{School of Physical Sciences}}\\
\small{\textit{National Institute of Science Education and Research, HBNI, Jatni - 752050, Odisha, India}}\\}

\begin{abstract}
In pursuit of a minimal basis for systems with non-ideal bond angles, 
in this work we try to pinpoint the exact orientation of the major 
overlapping orbitals along the nearest neighbouring coordination segments 
in a given systems such that they maximally represent the covalent interactions
through out the system.
We compute Mayer's bond order, akin to the Wiberg's bond index,
in the basis of atomic Wannier orbitals with 
customizable non-degenerate hybridization constructed from first principles,
in a representative variety of molecules and layered systems.
We put them in perspective with unbiased maximally localized descriptions of 
bonding and non-bonding orbitals, and energetics to tunneling of electrons through them
between nearest neighbours, to describe the different physical aspects of covalent interactions,
which are not necessarily represented by a single unique set of atomic or 
bonding orbitals.
\end{abstract}

\maketitle
\section{Introduction}
\label{intro}
To represent the electronic structure of a given covalent system with minimal tight-binding parameters
it is preferable to resort to a directed localized basis 
\cite{slater1931directed,kirkwood1977generalized,baxter1996molecular}
such that the basis orbitals maximally represent the dominant covalent interactions in the system. 
At a fundamental level the problem is essentially that of finding the orientation of the atomic orbitals
such that a minimum number of them facilitate maximum sharing or tunneling of electrons between neighbouring atoms.
An associated problem is to partition the electrons in a covalent system among atoms and bonds 
\cite{mayer2007bond}
such that the population of bonds are contributed by a minimum number of orbitals. %
Solutions are rather straightforward for systems with ideal bond-angles corresponding to degenerate hybridizations like $sp^{2,3}$. 
Complication arises with non-ideal bond angles, since for such systems, 
as we show in this work, making a choice becomes difficult, 
as different facets of covalent interactions are represented by different sets 
of orbitals for the same coordination. 
In fact, ``bent bonds'' \cite{wiberg1996bent} have been long suggested
in such systems, indicating deviation of orientation of atomic orbitals from
the direction of coordination, as they take part in such covalent bonds.

Hybrid orbitals have been central to description of covalent bonding
since their introduction\cite{pauling1931nature,slater1931directed} almost a century ago. 
%
Molecular orbitals theory based methodologies for construction of hybrid orbitals,
 \cite{del1963hybridization,mcweeny1968criteria,del1966bent,murrell1960construction,mayer1996atomic,mayer1995non}
 predating the advent of the Kohn-Sham(KS) density functional theory(DFT) \cite{hohenberg1964inhomogeneous,kohn1965self} 
based framework, have been grossly based on the maximum overlap condition, wherein
either the overlap matrix\cite{del1963hybridization,del1966bent}
or the first-order density matrices\cite{mayer1996atomic,foster1980natural},
calculated typically in the basis of  Slater \cite{slater1930atomic}
or the Gaussian \cite{hehre1969self,dunning1989gaussian} type orbitals,
are transformed into block diagonal forms, where each blocks are spanned by orbitals centered on a pair of
nearest neighbouring atoms.
%
%
The resultant variants of the hybrid orbitals
 like the  {\it natural hybrid orbitals}\cite{foster1980natural},
the {\it effective atomic orbital}\cite{mayer1995non},
the {\it generalized hybrid orbitals}\cite{baxter1996molecular,kirkwood1977generalized},
the {\it oriented quasi-atomic orbitals}\cite{west2013comprehensive},
or the ones constructed using the  {\it maximal orbital analysis}\cite{dupuis2019maximal} approach,
constituted the bedrock for understanding chemical bonding in molecules,
although limited or biased by the selection of the semi-analytic basis states with 
adjustable parameters.
%
%
%
With the advent of DFT \cite{kohn1965self} based computation of
electronic structure from first principles, 
attempts to construct localized description of electronic structure in the basis of the KS single particle states,
has been primarily undertaken in terms of the spatially localized Wannier functions(WF),\cite{wannier1937structure} 
which rendered  bonding and non-bonding orbitals if constructed from the occupied KS states.
Since WFs cannot be uniquely localized in more that one direction simultaneously unless facilitated by symmetry,
template based construction of WFs with numerically chosen gauge for the KS states to ensure maximal 
localization, \cite{marzari1997maximally,mostofi2008wannier90} has been the mainstay.
However with only $\Gamma$ point, as is the case for finite systems and acceptable for 
large super-cells,
the maximally localized WFs can be constructed without using any 
template\cite{mostofi2008wannier90,gygi2003computation,bhattacharjee2015activation}, 
as done in this work. 

%
Methodologically, in this work we introduce the notion of maximally valent orbitals (MVO), 
which are essentially a selection of major overlapping orbitals along coordinations, 
oriented such that a minimal
of them maximally account for sum of bond-orders along coordinations across the system, at the level of
nearest neighbourhoods or beyond.
Wannier function based on the template of MVOs thus constitute the maximally covalent Wannier functions(MCWF). 
We demonstrate search of MVOs within the sets of orthonormal Wannierized counterparts of non-degenerate \sptwo\ and \spthree\  
orbitals, referred here onwards as the n-\sptwo\ and n-\spthree\ orbitals, with customizable orientation, constructed from first principles.
The n-\sptwo\ and n-\spthree\ orbitals are the custom hybridized atomic orbitals(CHAO) with tunable hybridization as per
the geometry of nearest neighbour coordination around atoms in systems with non-ideal bond angles.
CHAOs are generalization of degenerate HAOs constructed from KS states of isolated atoms as demonstrated in 
Ref. \cite{hossain2021hybrid}. 
At the heart of the search of MVOs is the formulation and computation of bond-order in the basis of CHAWOs,
as defined by Mayer \cite{MAYER1985396} 
and found analogous to the Wiberg's bond index \cite{WIBERG19681083} priorly introduced.
We further calculate energetics and tight-binding parameters in the basis of CHAWOs,  
and calculate their projection on maximally localized WFs constructed without any template of CHAOs, 
in order to compare MVOs and MCWFs with other possible descriptions of atomic and bonding orbitals 
representing different aspects of covalent interactions,
demonstrated in a wide range of systems starting with cyclopropane which has the smallest C-C-C bond, 
to cyclobutadiene, diborane, ammonia and water,
and finally fullerene and some layered materials like silicene, germanene and MoS$_2$, all with bond angles different from that of 
degenerate \sptwo\ or \spthree\ coordination.
%

%
\section{Methodological details}
In this section first we briefly outline the construction of the template free maximally localized WFs used in this work,
and the HAOs, following similar approach.
Next we describe construction of CHAOs from HAOs and their Wannierization,
followed by formulation of bond-order in terms of the Wannierized CHAO, that is, the CHAWOs.
Finally we introduce MVOs as a particular choice of CHAWOs, and MCWOs.

The construction of the template free MLWFs\cite{gygi2003computation,bhattacharjee2015activation}
 to describe bonding and non-bonding orbitals, 
is precursor to the construction of HAOs \cite{hossain2021hybrid}, 
in terms of the technique for spatial localization. 
The only difference is that the MLWFs are constructed exclusively within the subspace of the occupied KS states 
of a given system, while the HAOs are constructed within an extended sub-space beyond the occupied sub-space of an isolated atom.
%
The localization scheme in both cases is based on maximal joint diagonalization of the generally 
non-commuting set of the first moment matrices(FMM) which are the representation of the three position operators 
$\hat{x},\hat{y},\hat{z}$ within a finite sub-space of basis states. 
The procedure 
follows from the fact that the total spread of a set of finite  ($N$) number of orbitals along $\hat{x}$, 
given by:
\begin{equation}
\Omega_x = \sum_ {i=1,N} \left[ \langle \phi_i | x^2 | \phi_i \rangle 
- |\langle \phi_i | x | \phi_i \rangle|^2 \right],
\end{equation}
can be expressed as:
\begin{eqnarray}
\Omega_x
&=& \sum_ {i=1,N}  \left( \sum^N_{j\ne i} |X_{ij}|^2 +   \sum^{\infty}_{j=N+1} |X_{ij}|^2 \right).
\label{omegaWF}
\end{eqnarray}
where
$X_{ij}=\langle \phi_i \mid x \mid \phi_j \rangle$. 
%
The off-diagonal elements of the FMM in the first term in the RHS of Eqn.\ref{omegaWF}
are simultaneously minimized through an iterative scheme based on the Jacobi method of matrix diagonalization, 
wherein the off-diagonal elements of a single or a commuting set of matrices are
set to zero through successive application of two dimensional rotation. 
In case of a  set of non-commuting matrices, a choice of rotation matrices 
which will maximally diagonalize the non-commuting matrices has been derived in Ref. \cite{cardoso1996jacobi}.  
The same has been used in this work, as well as for construction of HAOs described in Ref.\cite{hossain2021hybrid}, 
which may be refer for relevant details of computation of the rotation matrices.



%
Construction of CHAOs from HAOs involve two steps -
(1) Reconstruction of unhybridized atomic orbitals(UAO) from degenerate HAOs, and
(2) Re-hybridization of UAOs to construct CHAOs. 
%
In step 1, for a given element, linear combination of HAOs render UAO aligned perfectly 
as per a preferred Cartesian system of axes, 
with the variation of the radial part determined by the pseudo-potential used. 
%
In principle this process is straightforward since the analytic hybridization matrix for degenerate $sp^md^n$ hybridization is known. 
Surmountable technical complication arises with the arbitrary overall orientation of the set of degenerate HAOs.
%
%
%
Notably, up to $n=2$, UAOs obtained this way are essentially the rotated KS states, since for elements with
$2s$ and $2p$ valence electrons, the lowest three degenerate block of KS states are the three orthonormal 
$2p$ states in random orientation. 
However, for $n>2$ arbitrary mixing of degenerate KS states of the valence shells
makes it impossible to directly use them individually as pure atomic orbitals after simple rotation.
UAOs obtained from the HAOs, which are maximally localized by construction, are thus assured
to render the most localized form of pure orbitals aligned along any preferred set of Cartesian axes
as per the pseudo-potential used.  

Re-hybridization of UAOs to n-\sptwo\ or n-\spthree\ CHAOs 
are performed using hybridization matrices specific to symmetries as per that of the nearest neighbourhoods. 
%
For example, for CHAOs of the nitrogen atom in \ammonia\ 
we use a hybridization of form
\begin{equation}
\left( \begin{array}{cccc}
a & b & b & c \\
a & c & b & b \\
a & b & c & b \\
d & e & e & e \\
\end{array} \right)
\left( \begin{array}{c}
s \\
p_x \\
p_y \\
p_z \\
\end{array} \right).
\label{hmatone}
\end{equation}
The unitarity of the matrix allows one independent parameter and if we choose it to be $c$ then the
other parameters can be calculated as:  
$ a = \sqrt{-1-3c^2 + 4c}$, $b=c-1$, $e=-\frac{a}{\sqrt{(3c-2)^2+3a^2} }$, $d=\sqrt{1-3e^2}$.
%
%
More generally, an irregular tetrahedral orientation of orbitals can be assigned with hybridization matrix
of form:
\begin{equation}
\left( \begin{array}{cccc}
a & \frac{1}{\sqrt{2}} & \sqrt{\frac{1}{2}- a^2} & 0 \\
a &-\frac{1}{\sqrt{2}} & \sqrt{\frac{1}{2}- a^2} & 0 \\

b           \sqrt{\frac{1}{2}-a^2} & 0 & -ab & -\sqrt{1-\frac{b^2}{2}} \\
\sqrt{2-b^2}\sqrt{\frac{1}{2}-a^2} & 0 & -a\sqrt{2-b^2} & \frac{b}{\sqrt{2}} \\

\end{array} \right) 
\label{hmattwo}
\end{equation}
with two independent parameters $a$ and $b$ representing the two angles which complete the assignment
of four orthonormal orbitals.
As evident in the matrix, we consider two of the orbitals, the first two, oriented in xy plane
symmetrically about the y axis, while the other two orbitals are in the yz plane.
The third and fourth orbitals can also be chosen to be symmetric about the y axis, which reduces the number
of independent parameters to one, and geometrically akin to the majority of tetrahedral coordination,
like those of C in C$_n$H$_{2+2n}$.
%

Given a system of atoms, we construct separate sets of CHAOs for atoms
of different elements and relative orientation of nearest neighbours(nn) around them.
Through choice of parameters in the hybridization matrix we can orient the CHAOs exactly 
along the direction of coordinations, or in any systematic variation expressible in terms of those directions.
Sets of CHAOs  constructed for each such types of atoms are then transferred from their atomic nurseries 
to the given system and oriented according to nn coordinations around each atom, 
to constitute a set of localized non-orthogonal basis made of transferred CHAOs with intra-atomic orthogonality.
%

Wannierization of the transferred CHAOs, say $N$ in number, in the basis of $N_{KS}(\ge N)$ number of KS states, 
starts with construction of a set of quasi-Bloch states
$\left\{ \tilde{\psi}_{\vec{k},j}(\vec{r}) \right\}$  from CHAOs,  
and subsequently projecting them  
%
on the orthonormal Bloch states constructed from the KS single-particle states: 
\begin{equation}
 O(\vec{k})_{m,j}= \langle \psi^{KS}_{\vec{k},m}  \mid \tilde{\psi}_{\vec{k},j} \rangle.
\label{proj}
\end{equation}  
%
%
Overlaps between the non-orthogonal quasi-Bloch states are calculated within the manifold of the considered KS states as:
\begin{equation}
S(\vec{k})_{m,n}= \sum_l^{N_{KS}} O(\vec{k})^*_{l,m} O(\vec{k})_{l,n}.
\label{overlap}
\end{equation}  
%
Values of $\sum_{\vec{k}}|S(\vec{k})_{n,n}|^2/N_k$  implies representability of the $n$-th CHAO 
within the set of KS states considered, and should be typically above 0.85 for good agreement of KS band-gap
and valence band width, with those calculated with the resultant tight-binding parameters, in the covalent
systems made of $p$-block elements as mostly studies in this work.
%
%
%

Through L\"{o}wdin symmetric orthogonalization(LSO)\cite{lowdin1950non} a new set of orthonormal Bloch states
can be constructed as:
\begin{equation}
\Psi_{\vec{k},n}(\vec{r}) = \sum_m^N S(\vec{k})^{-\frac{1}{2}}_{m,n} \sum_l^{N_{KS}} O(\vec{k})_{l,m} \psi^{KS}_{\vec{k},l}(\vec{r}), 
\label{awobf}
\end{equation}
which can be used to construct an orthonormal set of localized Wannier functions 
referred in this paper as the custom  hybrid atomic Wannier orbitals (CHAWO):
%
\begin{equation}
\Phi_{\vec{R'},j}(\vec{r}) = \frac{1}{\sqrt{N_k}} \sum_{\vec{k}} 
 e^{-i \vec{k}\cdot \vec{R'}} \sum_l^{N_{KS}} U(\vec{k})_{lj} \psi^{KS}_{\vec{k},j}(\vec{r}).
\label{chawo}
\end{equation}  
where $U(\vec{k})=O(\vec{k})S(\vec{k})^{-\frac{1}{2}}$.
LSO chooses the appropriate linear combination of KS states such that
resultant CHAWOs are orthonormal yet substantially resemble the template of transferred CHAOs.
TB parameters in CHAWO basis is straightforwardly calculated as:
\begin{eqnarray}
& &t_{\vec{R'},\vec{R},i,j} = \langle \Phi_{\vec{R'},i} \mid H^{KS} \mid \Phi_{\vec{R},j} \rangle \nonumber \\
&=&\frac{1}{N_k}\sum_{\vec{k}}^{BZ}e^{i\vec{k}.(\vec{R'}-\vec{R})}\sum_l^{N_{KS}} U(\vec{k})^*_{li} U(\vec{k})_{lj} E^{KS}_{\vec{k},l}.
\label{hop}
\end{eqnarray}
where $\left\{E^{KS}_{\vec{k},l}\right\}$  are KS energy eigenvalues. 
With $N_{KS}>N$, $U(\vec{k})$ becomes semi-unitary, and spatial localization of CHAWOs enhances and eventually converges with $N_{KS}$.
However for this work we have restricted $N_{KS}=N$ so that $U(\vec{k})$ is square matrix whose inverse 
can be unambiguously invoked in order to expand KS states completely in terms of CHAWOs.
%

%
Notably,  representability of the UAOs or HAOs or CHAOs in the KS states of the given system where they are to be Wannierized,
can be maximized by choosing to constructed them using the same pseudo-potentials which are used to 
compute the KS states of the given system.
High degree of representability ensures consolidation of the $O$ matrix [Eqn.(\ref{proj})] over fewer bands of KS states, 
which in turn consolidates localization of the Wannierized orbitals.
In principle, for a given system  we could also 
directly Wannierize a template of analytic or semi-analytic  orbitals such as the hydrogenic, Slater or Gaussian type orbitals, 
or their hybrids, instead of the UAOs or the HAOs or CHAOs which are purely numerical in nature. 
However, unlike the numerical ones which can be chosen to have maximum representability 
by construction, the enhancement of representability of the analytic orbitals require numerical optimization of parameters 
used in defining those orbitals.
%
%
\subsection{Bond-order in CHAWO basis}
To derive an expression of bond-order(BO) similar to that proposed by Mayer \cite{MAYER1985396,mayer1986bond}
, we start with the traditional or classical definition of BO involving 
the $i$-th and $j$-th atomic orbitals for a given spin:
\[\mbox{B}_{\vec{R'}\vec{R},ij} = \frac{ n^{+}_{\vec{R'}\vec{R},ij}-n^{-}_{\vec{R'}\vec{R},ij}}{2}, \]
$n^{\pm}_{\vec{R'}\vec{R},ij}$ being the occupation of the bonding(+) and anti-bonding(-) orbitals considered in the CHAWO basis as:
\[ \phi^{\pm}_{\vec{R'}\vec{R},ij} = \frac{1}{\sqrt{2}}\left(\Phi_{\vec{R'},i} \pm \Phi_{\vec{R},j}  \right), \]
%
Within the subspace of occupied KS states:
\begin{eqnarray}
& &\mbox{B}_{\vec{R}'\vec{R},ij}
= \mbox{Re}[\langle \Phi_{\vec{R}',i} \mid \hat{P} \mid  \Phi_{\vec{R},j} \rangle ]\nonumber\\
&=& \sum_{\vec{k}}^{BZ} \sum_l^{N_{KS}} \frac{f_{\vec{k},l}}{N_k}
\mbox{Re}[ e^{i\vec{k}.(\vec{R}'-\vec{R})} U(\vec{k})^*_{li} U(\vec{k})_{lj} ]
\label{bo}
\end{eqnarray}
where  $\hat{P}$ is the projection operator for a given spin:
\[ \hat{P}=  \sum_{\vec{k}}^{BZ} \sum_l^{N_{KS}} \mid \psi^{KS}_{\vec{k},l} \rangle  f_{\vec{k},l} \langle \psi^{KS}_{\vec{k},l} \mid \]
$f_{\vec{k},l} $ being the occupancy of the $l$-th KS state with wave-vector $\vec{k}$.
$\mbox{B}_{\vec{R}'\vec{R},ij}$ in (\ref{bo}) is essentially the Coulson's bond order(CBO)\cite{coulson1939electronic}, 
used primarily in case of a single orbital per atom,  where $i$ and $j$ effectively become the atom indexes.
For a given covalent bond, CBO values, as evaluated in (\ref{bo}), can be positive or negative depending on the relative phase of the two orbitals involved.
This indicates that CBO values can not be associated with any form of electron population. 
In fact, the total number of electrons for a given spin:
\begin{eqnarray} 
N_e &=& \frac{1}{N_k}  \sum_{\vec{k}}^{BZ} \sum_l^{N_{KS}} 
 \langle \psi^{KS}_{\vec{k},l} \mid  \hat{P} \mid \psi^{KS}_{\vec{k},l} \rangle \nonumber \\
    &=& \frac{1}{N_k}  \sum_{\vec{R}} \sum_j^{N} 
 \langle \Phi_{\vec{R},j} \mid  \hat{P} \mid \Phi_{\vec{R},j} \rangle \nonumber \\
&=& \frac{1}{N_k}  \sum_{\vec{R}} \sum_j B_{\vec{R}\vec{R},jj} \nonumber\\
&=& \sum_A \sum_{j\in A} B_{00,jj}=\sum_A Q_A
\label{etot}
\end{eqnarray}
$Q_A$ being the number of electrons which can be associated with atom A.
Notably, $Q_A$ is analogous to the Mulliken's \textit{gross atomic population}\cite{mulliken1955electronic},
 which is same as the \textit{net atomic population} in case of orthonormal basis, like the CHAWOs, since the 
\textit{overlap population} vanishes due to the orthonormality of the basis in the Mulliken's population analysis scheme. 
Eqn.(\ref{etot}) also reiterates that the CBO values can not be used in partitioning of electrons into atoms or bonds
since they do not contribute to the total number of electrons,
whereas, the general classical notion of bond-order is that it is half the total number of electrons shared in a covalent bond including both the spins.

However, using the indempotency of $\hat{P}$ for integral occupancy of states for a given spin, we can write:
\begin{eqnarray} 
N_e &=& \frac{1}{N_k}  \sum_{\vec{k}}^{BZ} \sum_l^{N_{KS}} 
 \langle \psi^{KS}_{\vec{k},l} \mid  \hat{P}\hat{P} \mid \psi^{KS}_{\vec{k},l} \rangle \nonumber \\
 &=& \frac{1}{N_k}  \sum_{\vec{R}} \sum_j^{N} 
 \langle \Phi_{\vec{R},j} \mid  \hat{P}\hat{P} \mid \Phi_{\vec{R},j} \rangle 
\label{pp}
\end{eqnarray}
Inserting $ \sum_{\vec{R}} \sum_l^{N}  \mid \Phi_{\vec{R},l} \rangle\langle \Phi_{\vec{R},l} \mid $ between the two $\hat{P}$ in (\ref{pp})
we obtain:
\begin{eqnarray} 
N_e &=& \frac{1}{N_k} \sum_{\vec{R}} \sum_j^{N}  \sum_{\vec{R}'} \sum_l^{N} 
B_{\vec{R}\vec{R}',jl}B_{\vec{R}'\vec{R},lj}\nonumber\\
&=&  \sum_j^{N}  \sum_{\vec{R}'} \sum_l^{N} B_{0\vec{R}',jl}B_{\vec{R}'0,lj}
\label{wibergone}
\end{eqnarray}
using (\ref{bo}). In (\ref{wibergone}), for an  atom in the $0$-th unit-cell, all other atoms in the $0$-th or in any other unit-cell($\vec{R}'$)
can be generalized as neighbours.  Therefore we can generalize (\ref{wibergone}) and partition $N_e$ as:
\begin{eqnarray} 
& &N_e =  \sum_j^{N} \sum_l^{N\times N_k} B_{jl}B_{lj} \nonumber\\
&=& \sum_A^{N_A} \left[\sum_{j\in A} \sum_l^{N\times N_k} B_{jl}B_{lj}\right]\nonumber\\
&=&\sum_A^{N_A} \left[\sum_{j\in A}  \left[\sum_{l\in A}B_{jl}B_{lj}+\sum_{l\not\in A}B_{jl}B_{lj}\right] \right] \nonumber \\
&=&\sum_A^{N_A} \left[\sum_{j\in A} \sum_{l\in A}B_{jl}B_{lj}+  \sum_{j\in A}\left( \sum_{l\in B\ne A}B_{jl}B_{lj}\right)\right] \nonumber \\
&=&\sum_A^{N_A} \left[ Q_{AA} +   \sum_{B\ne A} Q_{AB}\right]
\label{wibertwo}
\end{eqnarray}
where $Q_{AA}$ is the \textit{net atomic population} of atom A and the $Q_{AB}$ is the \textit{overlap population}
between atoms A and B, as defined by Mayer\cite{mayer2007bond}. 
Note that this version of overlap population is different from the Mulliken's overlap population since 
the later is zero for orthonormal basis. 

The quantity  $Q_{AB}$ was originally introduced as a bond-index\cite{WIBERG19681083} 
and has been later interpreted as the bond order 
following Wiberg's original introduction of \textit{covalent bonding capacity} of a basis orbital, say $\Phi_j$, as:
\begin{equation}
b_j = 2 P_{jj} - P^2_{jj}
\end{equation} 
where $P=\sum_\sigma P^\sigma $. 
As easily seen, $b_j$ will be 1 if   $\Phi_j$ takes part in a bonding orbital, and 0 if it is one in a lone-pair.
Using the indempotency property of $\hat{P}$ again,
\[ b_j = P^2_{jj}\]
which implies
\[ b_j = \sum_{l\in A} P_{jl}P_{lj} + \sum_{l\not\in A} P_{jl}P_{lj}\]
with more than one orbitals per atom.
Therefore if $j\in A$ then for the atom A we can write using (\ref{bo}):
\[ \sum_{j\in A}b_j = Q_{AA} + \sum_{B\ne A} Q_{AB},\]
where 
\[ Q_{AB}=  \sum_{j\in A} \sum_{i\in B} q_{AB,ij},\]
with
\[ q_{AB,ij}= P_{ji}P_{ij}\mid_{i\in A,j\in B}.\]
Thus with more than one basis orbitals centred on A, the net \textit{covalent bonding capacity} or the valency $V_A$ of atom A 
is assessed after subtracting the intra-atomic term $Q_{AA}$ from the net $\sum_{j\in A} b_j$:
\begin{equation}
V_A=\sum_{j\in A}b_j - Q_{AA}= \sum_{B\ne A} Q_{AB}
\end{equation}
Which notionally identifies $Q_{AB}$ as the bond order in the classical sense of valency of an atom
in a covalent system.

\begin{figure*}[t]
\flushleft
\textbf{Cyclopropane:} \\
\centering
\includegraphics[scale=0.35]{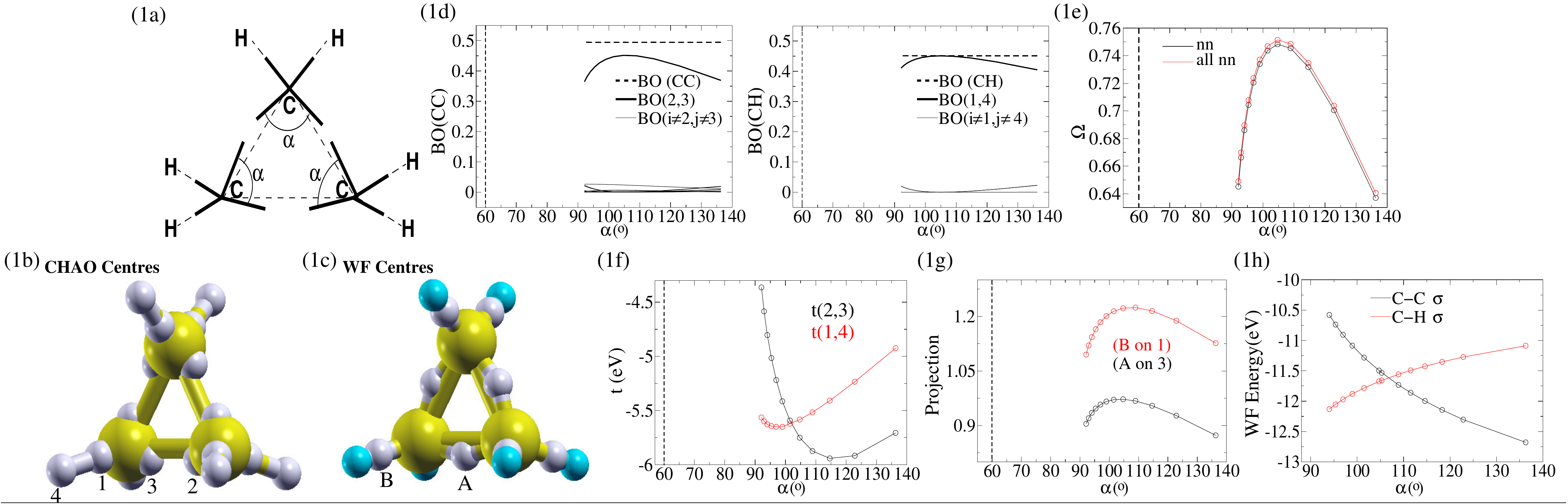}
\flushleft
\textbf{Cyclobutadiene:} \\
\centering
\includegraphics[scale=0.35]{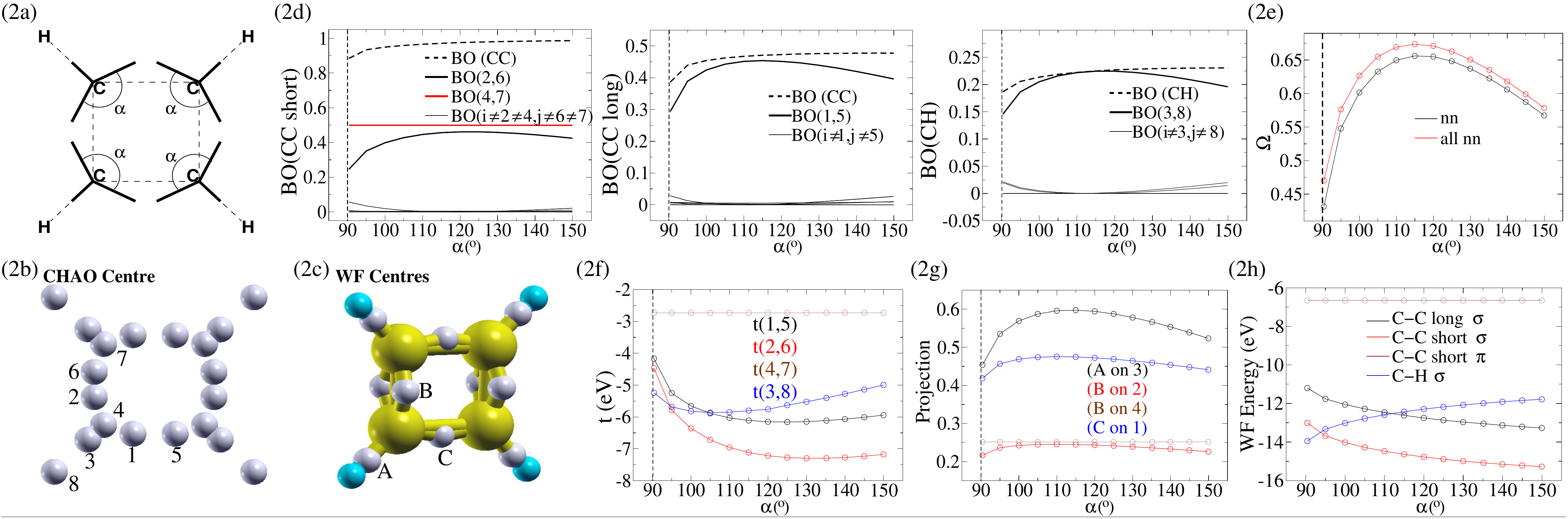}
\flushleft
\textbf{Diborane:} \\
\centering
\includegraphics[scale=0.35]{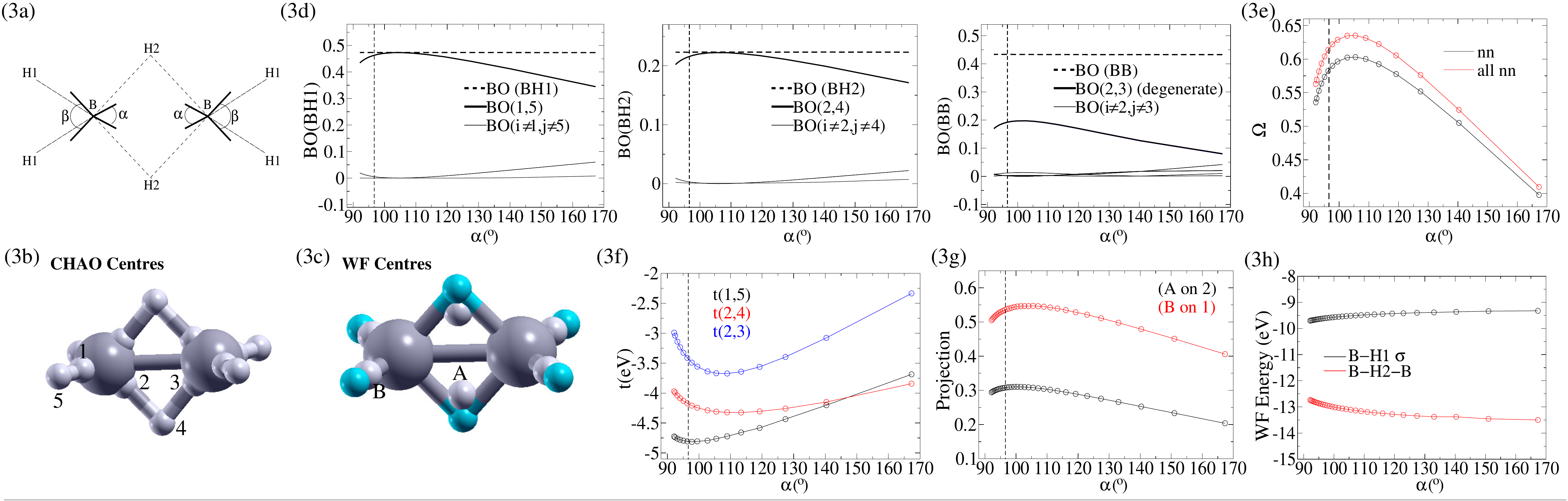}
\caption{
Plotted as function of relative angle $\alpha$[(a)] between CHAOs, 
variations of: 
(d) net BO and BO contributions from CHAWO pairs (whose charge centres are marked in (b)),
(e) $\Omega$[Eqn.(\ref{omega})],
(f) hopping parameter between the major overlapping CHAWOs,
(g) projection of template free WFs (whose charge centres are marked as in (c)) on CHAWOs marked in (b), 
(h) energetics of WFs made with template of CHAOs. 
}
\label{cyclopropane}
\end{figure*}

In this work we calculated bond orders as defined by $Q_{AB}$ and their decomposition
in orbital pairs, as a function of orientation of the CHAWOs, in order to pinpoint the orientation 
which maximises the BO contribution from the dominant orbital pairs for a given pair of nearest neighbouring atoms.
Since the net BO remains largely constant over different orientations [Fig.\ref{cyclopropane}] the dominant contribution can be numerically
traced as the maxima of the variance of contributions from different pairs for a given coordination.  
%
%
In fact, the set of CHAWOs which maximizes the sum of standard deviation of BO contributions 
of all the coordinations in a given system should in principle pinpoint to the orientation
of CHAOs which would render CHAWOs such that a minimum of them  would maximally
incorporate covalent interaction along all coordinations, led by the nn coordinations.
We therefore propose to seek the maxima of: 
\begin{equation}
\Omega=\sum_A \sum_{B\ne A} \sum_{i\in A} \sum_{j\in B} \frac{(\bar{q}_{AB} - q_{AB,ij})^2}{n_A n_B},
\label{omega}
\end{equation}
where
\[ \bar{q}_{AB} = \sum_{i\in A} \sum_{j\in B\ne A} \frac{q_{AB,ij}}{n_A n_B}, \]
$n_A$ and $n_B$ being number of orbitals centered on atoms A and B respectively. 
%
%
CHAOs corresponding to the maxima of $\Omega$ can thus be referred as \textit{maximally valent hybrid atomic orbitals}(MVHAO).
Correspondingly, 
WFs constructed using the template of MVHAOs can thus be referred as the  \textit{maximally covalent Wannier functions}(MCWF).
Notably, for systems with inequivalent atoms, finding $\Omega$ is in principle a multi-variable maximization problem.
In this work however we have restricted to systems where single variable maximization of $\Omega$ is sufficient or effectively so. 
%
%
\section{Computational details}
All the ground state geometries as well as ground state electronic structures are calculated
using the Quantum Espresso (QE) code\cite{giannozzi2009quantum} which is a plane wave based 
implementation of density functional theory (DFT)\cite{hohenberg1964inhomogeneous,kohn1965self}. 
The BFGS scheme has been used to obtain the relaxed structures within the pseudo-potential used. 
For periodic systems variable cell relaxation has been performed to optimize lattice parameters and ionic positions. 
%
%
The KS ground states are calculated within the Perdew-Burke-Ernzerhof (PBE)\cite{gga2} 
approximation of the exchange-correlation functional.  
%
Plane wave basis with kinetic energy cutoff of of 60 Rydberg has been used for 
all systems considered in this work. 
We used a 21x21x1 Monkhorst-Pack grid of k-points for the layered systems. 
%


For construction of WFs, CHAWOs and calculation of TB parameters and BO, 
we use our in-house implementation 
which used the KS states computed by the QE code.
%
Towards construction of n-\sptwo\ and n-\spthree\ CHAOs, 
the \spthree\ HAOs for $n=2$ are constructed in this work  for B, C, N, O, S, Si, Ge and Sn  
using the lowest four KS states which include a triply degenerate block. 
For $n>2$, \spthreedfive\ HAOs are constructed for Mo atoms in the basis of the lowest four KS states and further five states 
starting from the 6th to the 10th, which divide in two degenerate groups made of two and three KS sates with small difference in eigenvalues, 
the 5th non-degenerate KS state being the 5$s$ state transferred as is.

\begin{figure*}[t]
\flushleft
\textbf{Water:} \\
\centering
\includegraphics[scale=0.37]{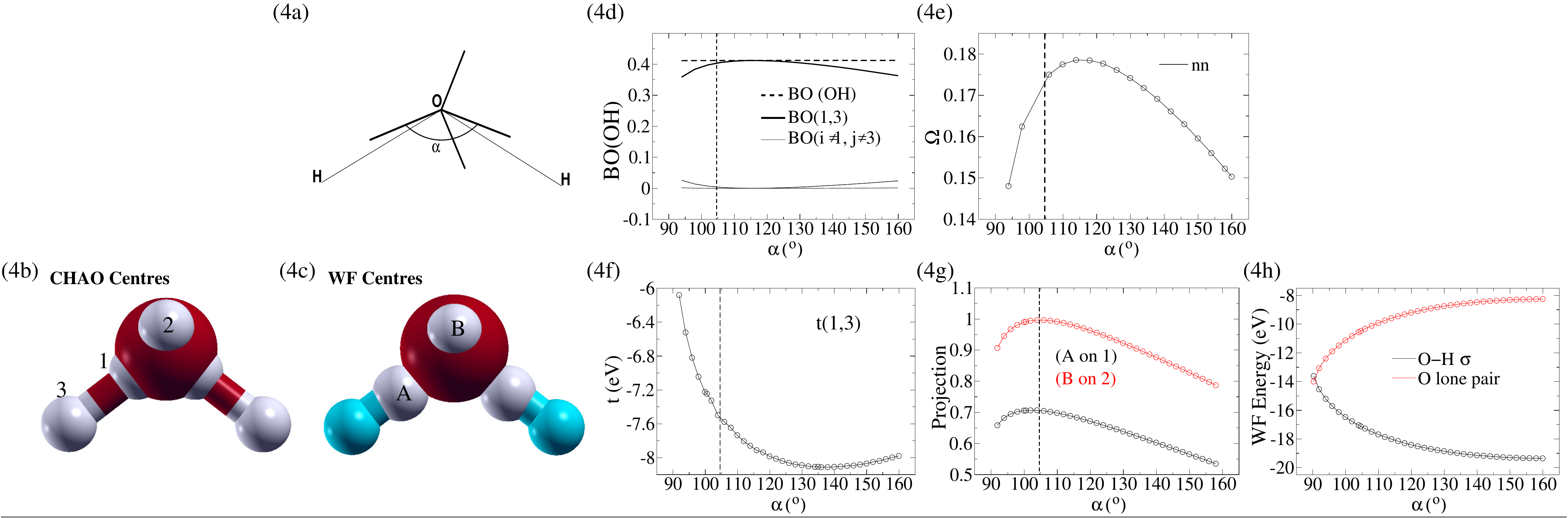}
\flushleft
\textbf{Ammonia:} \\
\centering
\includegraphics[scale=0.37]{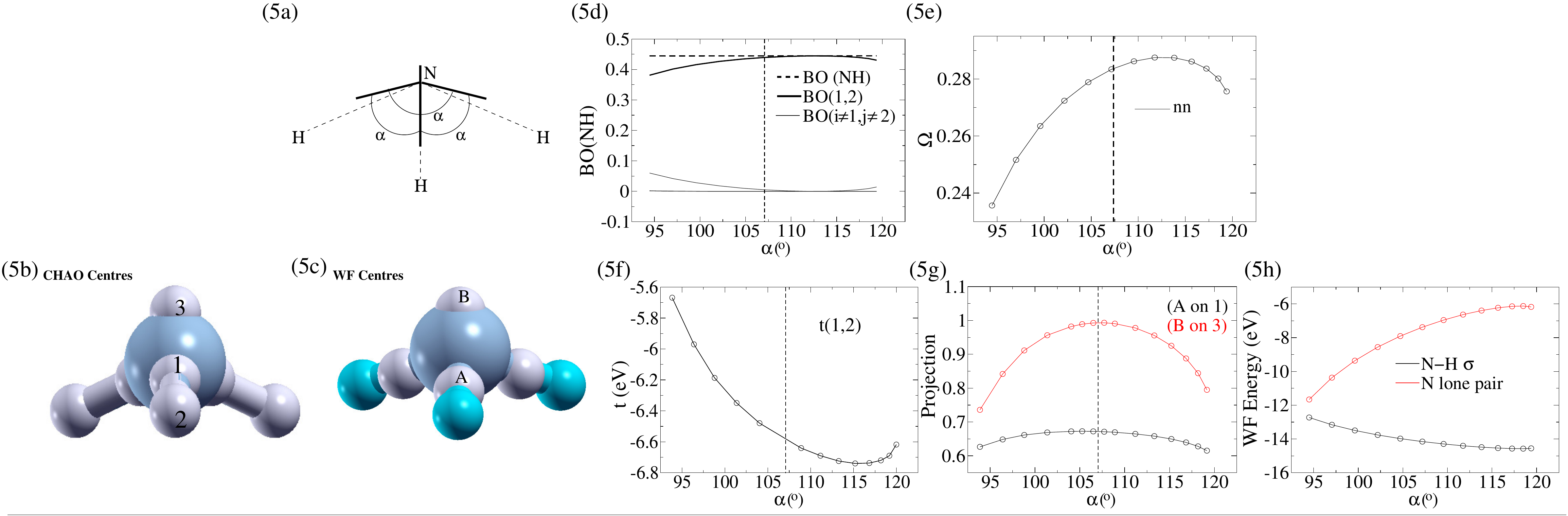}
\caption{Similar to Fig.\ref{cyclopropane},
Plotted as function of $\alpha$[(a)], 
variations of: 
(d) net BO and BO contributions from CHAWO pairs marked in (b),
(e) $\Omega$,
(f) hopping parameter between the major overlapping CHAWOs,
(g) projection of template free WFs marked as in (c), on CHAWOs marked in (b), 
(h) energetics of WFs made with template of CHAOs.
}
\label{water}
\end{figure*}
\section{Results and discussion}
\label{oo}
%
To choose the criteria of seeking optimal orientation 
of CHAOs as they take part in covalent bonding,  
we can in principle take recourse to the different descriptions of covalent bond based 
on different physical aspects of covalent interaction, for the same coordination.
Different choices of bonding orbital along a nn coordination,
can differ on the degree of sharing of electrons between atoms,
or the degree of spatial localization of electrons participating in the bond,  
besides the energetics of the orbitals.
Similarly, different choices of orientations of CHAOs would differ
not only on the  degree of sharing of electrons they facilitate between atoms,
but also on energetics of hopping of electrons through them, which has bearing on the strength of the covalent interaction
they would support. 
%
%
Since systems with non-ideal bond angles have been reported to have bent bonds\cite{wiberg1996bent}, 
making a choice of an optimal orientation would also thus amount to 
substantiating the bent nature from the deferent perspectives of covalent interaction.
For a given nn coordination, we therefore first look for
 the peak for $\Omega$ [Eqn.\ref{omega}]to find the orientation of the CHAOs which maximally hosts the covalent interactions as described by BO.
%
Secondly, the maxima of projection of the template free maximally localized bonding WFs on CHAWOs,
with the aim of seeking the CHAOs which would lead to the most localized description of the covalent interaction 
pointing arguably to the shortest path of tunnelling of electrons. 
%
%
%
And thirdly the  maxima of the magnitude of the hopping parameter($t$) between the two major overlapping CHAWOs
to find the energetically most favourable route of tunnelling of electron for the given coordination.
Symmetry of cyclopropane allows a single variable maximization of \omeg\ with one independent parameter
in the hybridization matrix.
As evident in Fig.\ref{cyclopropane}(1d), 
the relative angle \alfabo, which is the \alfa\ where the dominant BO contribution maximize for a given coordination,
are close for C-C and C-H coordinations, and thus coincide with \alfaomega\ which is 
where the \omeg\ [(1e)] maximizes. 
%
The value of \alfawf, which is the \alfa\ where the dominant projection of the template free MLWF on CHAWOs 
maximize[Fig.\ref{cyclopropane}(1g)], for both the coordinations, is also close to \alfaomega\  - at around 105\dg.
Thus the CHAWOs of C at \alfaomega, namely, the  MVHAOs of C as defined above, 
as well as the CHAWOs of C which have maximum overlap with the MLWFs representing the bonds, 
both have similar orientation and deviate from the  directions of C-C and C-H  by about 22.5\dg\ 
and 0.4\dg\ respectively.
Therefore for cyclopropane both the kinds of bonding orbitals - MLWFs as well as MCWFs,
are essentially same. 
Interestingly, while the values of \alfabo\ and \alfawf\ coincides with \alfaomega, 
the values of \alfat, which is the \alfa\ where $|t|$ maximizes[Fig.\ref{cyclopropane}(1f)] for the major overlapping CHAWOs, 
occur respectively at lower and higher angles than \alfaomega\ for the C-H and C-C coordinations.
Such a trends of \alfat\ values is consistent with the fact that the energies of the  C-H$\sigma$ and C-C$\sigma$ WFs 
constructed based on template made of CHAOs, show a crossing[Fig.\ref{cyclopropane}(1h)] around  \alfaomega, 
with the C-C(C-H)$\sigma$ being lower in energy below(above) \alfaomega.
These trends 
clearly suggest a competing 
preference of the two bonds, to deviation or ``bending'' from their respective directions of coordinations, above and below \alfaomega.
%
%
\begin{figure}[t]
\includegraphics[scale=0.44]{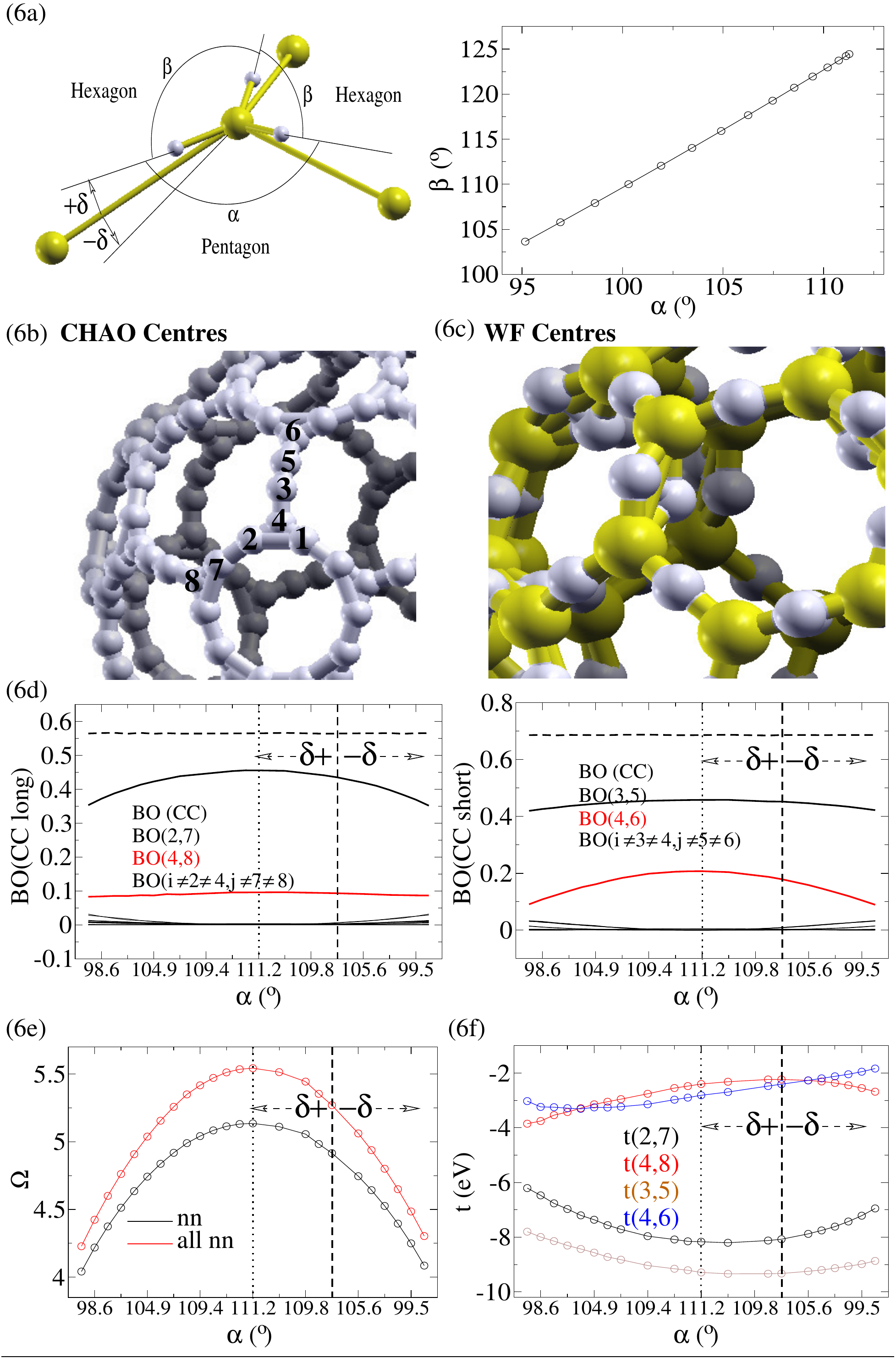}
\caption{{\bf{C$_{60}$}}: Plotted as function of $\alpha$ as shown in (6a) -
(6d) net BO and BO contributions from the Wannierized pair of CHAOs marked in (6b),
(6e) $\Omega$,
(6f) hopping parameter between the major overlapping CHAWOs.
Charge centres of template free MLWFs are shown in (6c).
Note that same values of \alfa\ recurs about the dotted line which corresponds to co-planarity of
$\alpha$ and $\beta$ : $\alpha+2\beta=360^\circ$.
The ($\alpha,\beta$) trajectory considered is plotted in 6a.
}
\label{fullerene}
\end{figure}

In case of cyclobutadiene, the optimization of CHAWOs of C atoms is essentially a problem of
two variable maximization of 
$\Omega$ owing to the lack of symmetry about C due to the inequivalent C-C bonds.
However, motivated by the small deviation of MVHAOs from the C-H coordination in cyclopropane,
we limit optimization of CHAOs for cyclobutadiene to their symmetric orientations about the C-H coordination, 
as evident in  Fig.\ref{cyclopropane}(2a).
%
%
%
%
Within such a constraint, the dominant BO contributors for the C-H  and the longer 
C-C coordination occurs between 110\dg\ and  115\dg\ [Fig.\ref{cyclopropane}(2d)], 
while it is about 120\dg\ for the  shorter C-C coordination,
leading to an  \alfaomega\ arond 115\dg,
implying a deviation of MVHAOs from both the shorter and longer C-C coordinations 
by about 12\dg. 
%
%
Similar to cyclopropane, 
firstly, the values of \alfawf\ [Fig.\ref{cyclopropane}(2g)] 
in this case are also close to the respective values of \alfabo, and secondly,
the values of \alfat\ [Fig.\ref{cyclopropane}(2f)] for C-H \sig\ and both the C-C\sig\ bonds, 
are respectively lower and higher than the 
\alfaomega\ value. 
Competing preference to deviation from direction of coordination, as seen in cyclopropane, is also evident in cyclobutadiene 
from the variation of energetics of the 
C-H \sig\ and the C-C\sig\ bonds [Fig.\ref{cyclopropane}(2h)] with \alfa. 
%
%
%
%

Next we calculate CHAWOs for diborane molecule well known for the B-H-B three centre two electron bond,
which in this work is marked by the BO values of B-B and B-H2 [Fig.\ref{cyclopropane}(3f)] of about 0.425 and 0.225 per spin, implying a total of 
about $2\times (\frac{0.425}{2}+2\times 0.225)=1.325$ electrons per spin for each of the 
B-H-B bond.
Symmetry of diborane allows single parameter maximization of \omeg\ .
Values of \alfabo\ [Fig.\ref{cyclopropane}((3d))], although are generally close to \alfazero, are larger for the B-H1 and B-H2 coordinations than that for the B-B coordination, 
leading to an intermediate value of \alfaomega\ [(3e)] close to \alfazero,
implying a deviation of MVHAOs by about 4\dg\ from both the B-H1 and B-H2 coordinations. 
Deviation of the \alfawf\ values[Fig.\ref{cyclopropane}(3g)] from \alfazero\ for the B-H2-B and B-H1 WFs are similar to those of the
respective \alfabo\ values.
Thus in this case also the MLWFs and the MCWFs suggest similar deviations of the   B-H2-B and B-H1 bonds from the directions of
B-H2 and B-H1 coordinations.
Among the hopping parameters, like we saw in case of cyclopropane and cyclobutadiene, 
the deviation of \alfat\ [Fig.\ref{cyclopropane}(3f)] of the B-H1, which is B-H \sig\ in this case, is the least,
which is also corroborated by the energetics[(3h)] of the template based WFs,
as it shows that the  B-H1 bonds do not prefer deviation,
while the B-H2-B three centre bonds do. 
Thus the competing energetics of B-B covalent interaction mediated by the H, 
and that of the C-H \sig\ bonding, with the latter dominating over the former due to multiplicity,
determines the structure of diborane. 

For molecules of water and ammonia[Fig.\ref{water}], each with only one kind of coordination, 
\alfabo\ [Fig.\ref{water}(4d,5d)] and \alfaomega\ [(4e,5e)] are same, and  
are clearly higher than their corresponding  \alfawf\ values [(4g,5g)],
which interestingly almost coincides with the actual bond angles \alfazero\ for both the molecules.
Therefore, while the template free MLWFs suggest no deviation in effect,
the MCWFs would suggest deviations of about 10\dg\ and 5\dg\ respectively from directions of 
of O-H and N-H coordinations.
Values of \alfat\ [Fig.\ref{water}(4f,5f)]for both the O-H and N-H \sig\ bonds clearly suggest even larger deviations,
which is consistent with the energetics[Fig.\ref{water}(4h,5h)] of the template based WFs for both the molecules,
since in both cases the bonding WFs energetically prefers deviation which is opposed by the lone pairs.
%

%
In \csixty, the two different C-C bonds - the shorter ones shared by two adjacent hexagons,
and the longer ones shared by hexagons with adjacent pentagons, demand a two parameter 
  maximization of \omeg\ in terms of \alfa\ and $\beta$ [Fig.\ref{fullerene}(6a)].
In this work however we restrict effectively to a one parameter optimization 
by seeking maxima of \omeg\ along the ($\alpha,\beta$) trajectory plotted in 
Fig.\ref{fullerene}(6a), which nevertheless
brings out the key aspect about the true nature of the MVHAOs.
For both the C-C bonds the \alfabo\ [Fig.\ref{fullerene}(6d)] and \alfaomega\ [(6e)] occur at around 111\dg\ where the 
three n-\spthree\ orbitals in effect become co-planar n-\sptwo,  
and the fourth one becomes pure \pz, 
implying a +$\delta$ deviation of MVHAOs
by about 13\dg\ and 11\dg\ from the direction of longer and 
shorter C-C coordination respectively.
%
%
Although the MLWFs [Fig.\ref{fullerene}(6c)] in this case renders the \pai\ bond exclusively along the shorter C-C coordination since the longer C-C bonds make pentagons,
comparable BO contribution exists between \pz\ orbitals 
along the longer C-C [(6d)] coordination as well. 
In fact the BO values along the two C-C coordinations are much comparable, in exception to that implied by MLWFs.
%
%
Notably, for both the C-C, \alfat\ for $t(2,7)$  and $t(5,3)$ 
suggests strongest \sig\-bond with similar or marginally less +$\delta$ deviation of participating CHOAs
from the C-C coordinations, compared to that implied by \alfaomega.
However, for the shorter C-C, the value of \alfat\ [Fig.\ref{fullerene}(6f)] for $t(4,6)$
suggests stronger \pai\-bond due to much larger +$\delta$ deviation
from C-C coordination than that implied by \alfaomega, 
which would push the major lobe of the unpaired n-\spthree\ orbital inside the fullerene cage. 
Thus the \alfaomega\ allows strong enough  \sig\-bonds but a weaker \pai-bond,
reiterating that it is primarily the C-C \sig\-bonds constituting the pentagons
which are responsible for the curved nature surfaces made of three coordinated carbon atoms 
with pentagon surrounded by hexagons. 
\begin{figure}[t]
\includegraphics[scale=0.67]{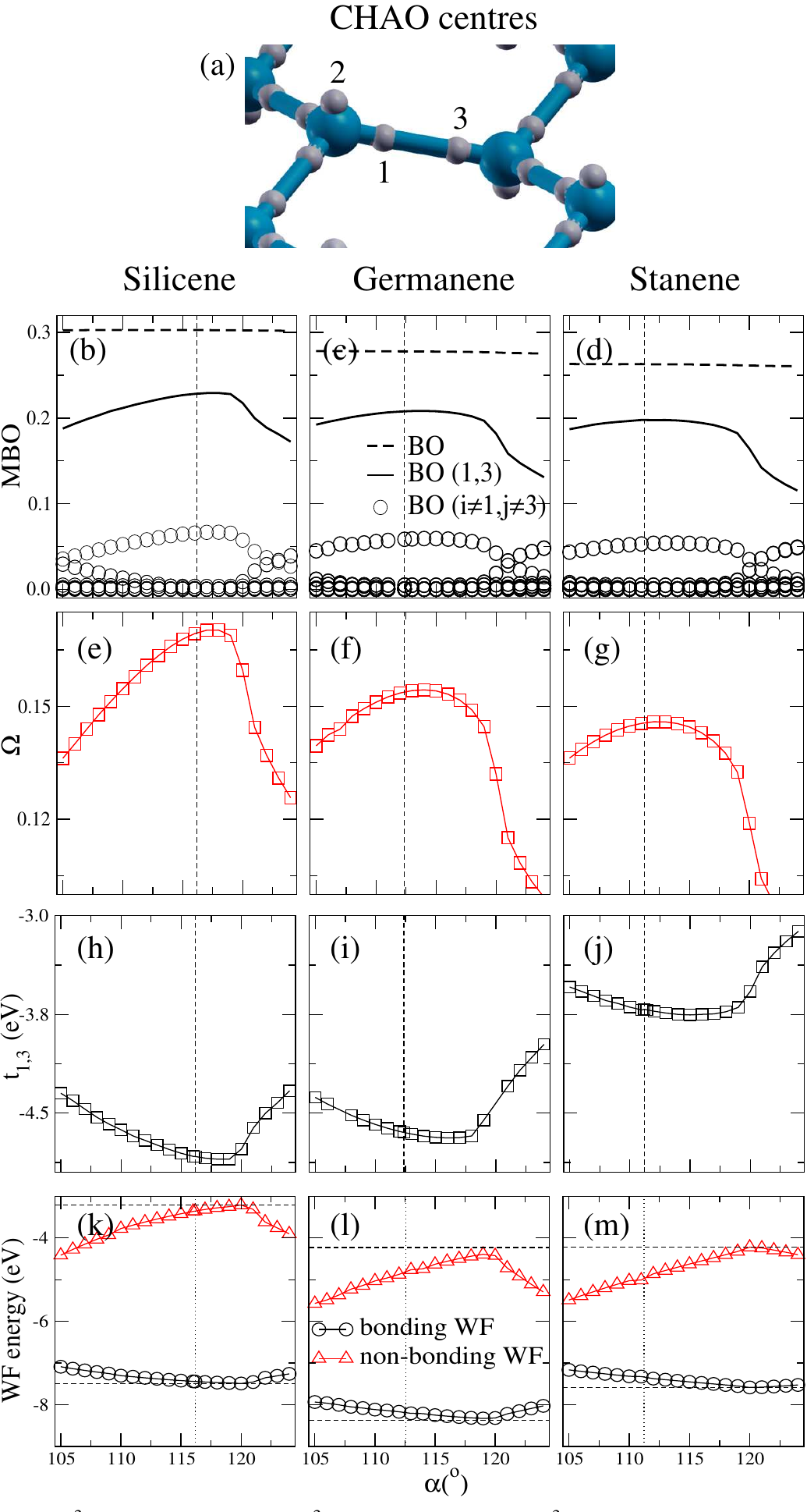}
\caption{
Plotted as a function of \alfa\ as shown in Fig.\ref{water}(5a) -
(b-d) total BO and contributions from orbital pairs marked in (a), 
(e-g) $\Omega$,
(h-j) $t$ between nn \sig\ ,
(k-m) Energy expectation values of WFs constructed using the template of CHAOs marked in (a).
}
\label{cor2d}
\end{figure}

In case of the layered systems formed from group 14 elements, moving down the group, 
the planar nature seen in graphene evolves into buckled structures of silicene, germanene and stanene with increasing non-co-planarity
marked by decreasing bond-angles - 
around 116.2$^{\circ}$, 112.3$^{\circ}$ and 111.2$^{\circ}$ respectively,  as per the pseudo-potentials used.
%
%
%
In search of MVHAO, we consider n-\spthree\ CHOAs, as considered for \ammonia\ .
With a consistent drop of the net nn BO[Fig.\ref{cor2d}(c-e)] from silicene  to stanene,
the dominant contributions to nn BO, 
appear to have a broadening peak marginally above \alfazero\ leading to
 \alfaomega\ [Fig.\ref{cor2d}(e-g)] close to \alfazero\ as well.
The deviation of MVHAOs are around 3\dg, 2\dg, and 0.6\dg\ for silicene,
germanene and stanene, in keeping with their increasing non-planarity.  
However the broadening peak of \omeg\ from silecene to stanene also suggest increasing lack
of directional preference for the CHAOs, consistent with decreasing BO,
corroborated by decreasing strength of covalent interaction reflected by
the values of hopping parameter[Fig.\ref{cor2d}(h-j)].
Interestingly, the energetics[Fig.\ref{cor2d}(k-m)] of WFs constructed using the template of CHAOs
suggests a strong preference for the bonding WFs to be  participated
by almost co-planar \sptwo\ orbitals at \alfa=120\dg,
which however, according to the energetics, is  least preferred by the non-bonding WFs.
%
%
%
%
These trends appear to suggest that lowering of kinetic energy due to delocalization of the unpaired n-\spthree orbital, is dominated by the energetics of the steric repulsion of the other three n-\spthree orbitals 
which are effectively co-planar as they take part in \sig\-bonds described by the WF, 
resulting into the increased degree of bucking with increasing Z.
%

\begin{figure}[b]
\includegraphics[scale=0.3]{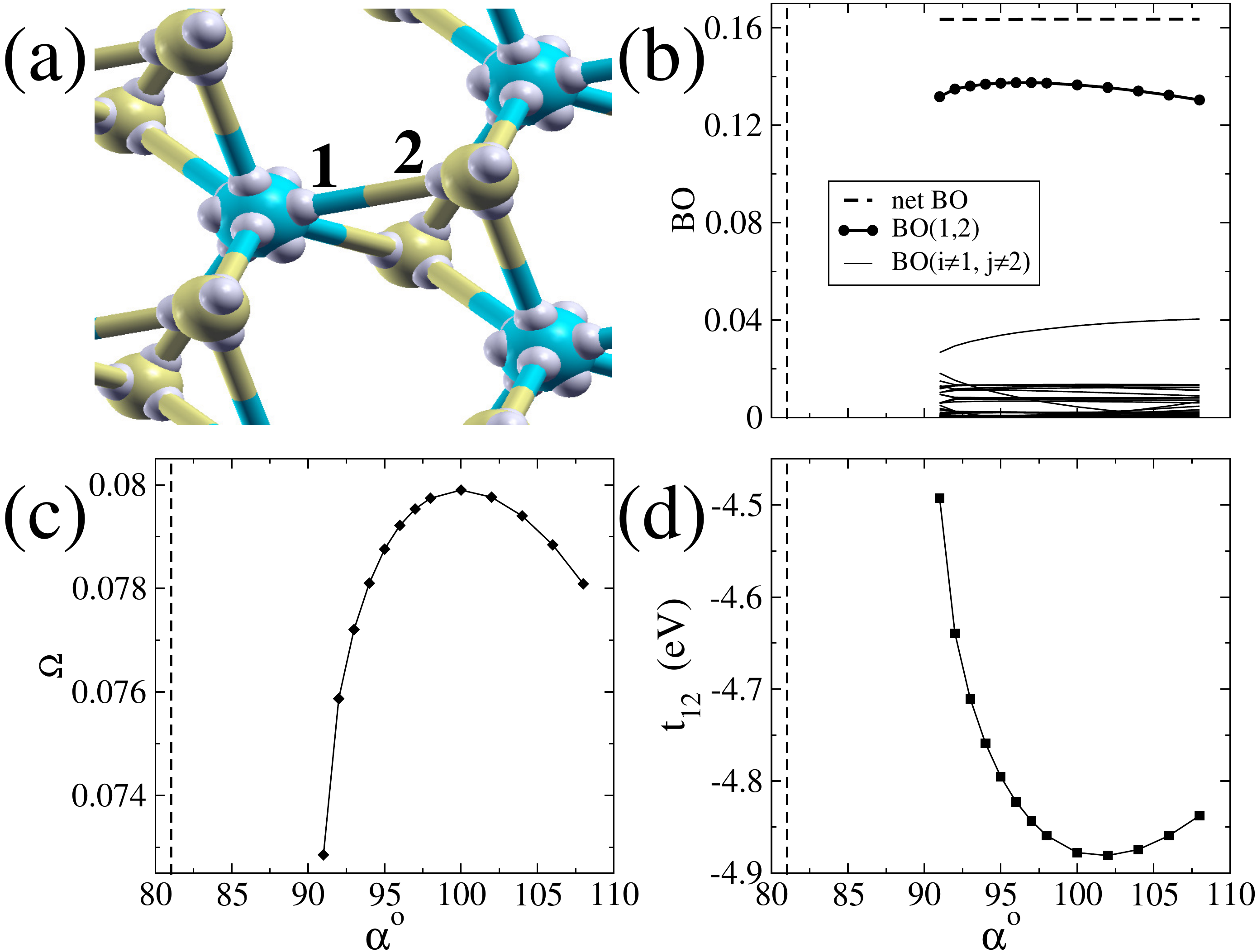}
\caption{
Plotted as a function of \alfa\ as shown in Fig.\ref{water}(5a) -
(b) total BO and contributions from orbital pairs marked in unitcell of MoS$_2$ (a), 
(c) $\Omega$,
(d) $t$ between  orbital pairs marked in (a). 
The dashed line corresponds to the actual bond angle (\alfazero).
}
\label{mos2}
\end{figure}
\begin{figure}[t]
\includegraphics[scale=0.35]{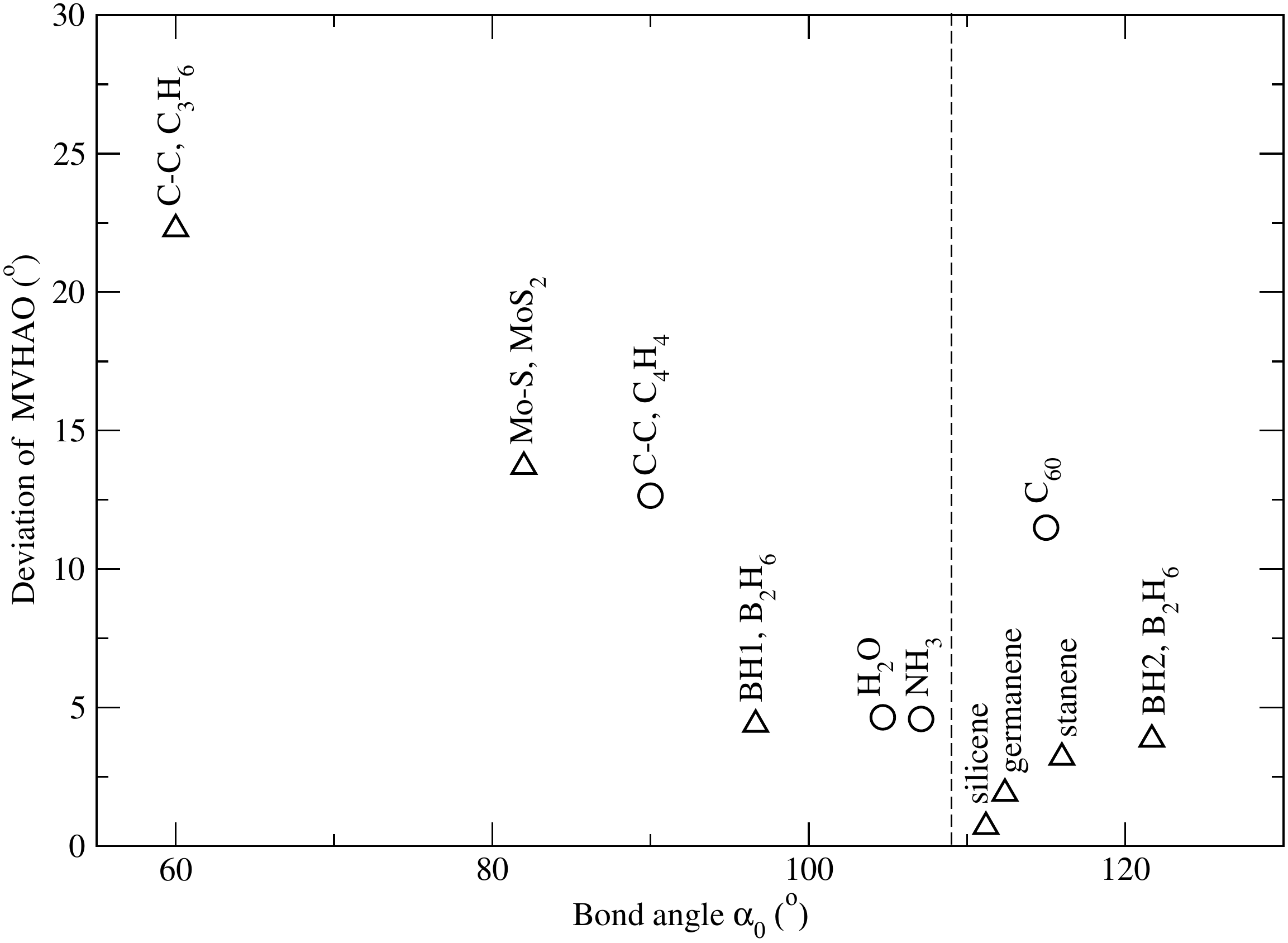}
\caption{
Plotted as a function of \alfazero, the deviation  of the MVHAOs from the coordination segment mentioned
above the data points. The dashed line is at the ideal bond-angle for tetrahedral coordination.
}
\label{devplot}
\end{figure}

In case of  molybdenum disulfide (MoS$_2$)[Fig.\ref{mos2}] monolayer, where the Mo-S-Mo angle is around 81$^{\circ}$,
we considered 5$s$ and 4$sp^3d^5$ HAWOS  [Fig.\ref{mos2}(a)] for Mo, 
as per the grouping of energetics of the KS states of an isolated Mo atom.
The  4$sp^3d^5$ hybridization results into three planar orbitals with C3 symmetry and six orbitals in 
trigonal prismatic orientation with symmetry as per the nearest neighbourhood of S atoms. 
For S,  we consider n-\spthree\  CHOAs, as we considered for \ammonia\ and the group 14 layered systems.
Since the orientations of the six trigonal prismatic orbitals of Mo can not change unless the planarity of the
other three trigonal orbitals is disturbed, we keep the orbitals of Mo unchanged and symmetrically vary 
the orientation of the n-\spthree\ CHAOs of S in search of \alfaomega, which we find
to be around 100\dg\ [Fig.\ref{mos2}(c)] which is close to \alfat\ [Fig.\ref{mos2}(d)] as well.
Thus the n-\spthree\ CHAOs of S participating in coordination with Mo make 100\dg\ among each other
and deviate by 13.7\dg\ from the direction of Mo-S coordination.
For a given Mo-S coordination, BO for both spins together is about 0.32 [Fig.\ref{mos2}(b)], 
while for an Mo-Mo coordination it is about 0.12.
The net atomic population ($Q_{AA}$) of Mo and S are about 11 and 4.7, implying a total of about 
20.5 electrons localized on atoms, out of the total of 26 electrons per unit-cell, 
due to 14(6) valence electrons of Mo(S) as per the pseudo-potential considered.  
Of the rest of about 5.5 electrons, the majority (0.32$\times$2$\times$6) is contribute by the six Mo-S bonds,
followed by the three Mo-Mo next nearest coordinations per unit-cell.
%
%

The deviation of MVHAOs from the direction of coordinations found in all the systems 
studied in this work is summerized in Fig.\ref{devplot},
where the deviations of MVHAOs centred on four(three)-coordinated sites, are shown by up-traingles(circles). 
Deviations of such MVHAOs increases in effect linearly with the degree of lowering of bond angle from the  
ideal tetrhedral bond-angle, with clear pinch-off at the ideal bond-angle.
Deviations of MVHAOs centred on three coordinated sites also appears to approache 
pinch-off at the ideal tetrhedral bond-angle with increasing 
tetrahedrality of the three nn coordinations and the lone-pair.
Substantial deviations in case of cyclobutadiene and \csixty\ 
are rooted at substantial differences among their bond-angles. 
As obvious, the effectiveness of MVHAOs as minimal basis increases with their increasing deviation
from nn coordinations.

\section{Conclusion}
%
In search of an optimally directed basis, 
we begin this work with construction of non-degenerate \textit{custom hybridized atomic orbitals}(CHAO) 
with variable orientation, from the degenerate set of hybridized atomic orbitals, 
in the basis of KS states of isolated atoms.
We next formulate Mayer's bond order in the basis of the Wannierized counterparts of the CHAO, 
constructed from the KS states of a given system, 
and introduce the \textit{ maximally valent hybrid atomic orbitals} (MVHAO) and the corresponding template based
WFs as the \textit{ maximally covalent Wannier functions} (MCWF), and use them to substantiate the
deviation of hybrid atomic orbitals from directions of coordinations as they participate in 
covalent bonding, as summerized in Fig.\ref{devplot}, leading to the bent nature of such bonds, 
in a host of molecules and layered systems with non-ideal bond angles.
Through comparison of bond-order(BO) contributions and hopping parameters from the Wannierized
 pairs of CHAOs,
and their overlap with template free maximally localized Wannier functions(MLWF),
we point out how maximally covalent representation of a given coordination can differ 
from its maximally localized, and energetically favourable representations
of covalent interactions in these systems,
shedding light on different perspectives of inter-atomic sharing of electrons in general. 
%
%
\section{Acknowledgements}
Computations have been performed in computing clusters supported by the 
Dept. of Atomic Energy of the Govt. of India.
\bibliographystyle{apsrev4-1}
\bibliography{references}


\end{document}